\documentclass{article}

\usepackage{spconf,amsmath,amssymb,amsbsy,amsxtra,graphicx}
\usepackage{amsmath,epsfig,amssymb,mathrsfs,psfrag,epsf,enumerate,bm,color,cite}

\ninept
\title{A Compressive Method for Centralized PSD Map Construction}
\name{Mohammad Eslami, Farah Torkamani-Azar, Esfandiar Mehrshahi}
\address{Cognitive Communication Research Group, \\Department of Electrical Engineering, \\ Shahid Beheshti University, Tehran, Iran}
\begin{document}
%
\maketitle
\small    

\begin{abstract}
Spectrum resources are facing huge demands and cognitive radio (CR) can improve the spectrum utilization. Recently, power spectral density (PSD) map is defined to enable the CR to reuse the frequency resources regarding to the area. For this reason, the sensed PSDs are fused by a Fusion Center (FC) which the sensed PSDs are collected by the distributed sensors in the area. But, for a given zone, the sensed PSD by neighbor CR sensors may contain a shared common component for a while. This component can be exploited in the theory of the distributed source coding (DSC) to compress sensing data more. In this paper based on the distributed compressive sensing (DCS) a method is proposed to compress and reconstruct the PSDs of the sensors when the data transmission is slightly imperfect. Simulation results show the advantages of using proposed method in compressing, reducing overhead and also recovering PSDs. 
\end{abstract} 


\section{Introduction and System Model}\label{Introduction}

In compressive spectrum sensing \cite{CrCs3}, the sensed PSD can be performed by $\boldsymbol s = \boldsymbol W \boldsymbol F \boldsymbol a$ where an smoothing operator $\boldsymbol W$ followed by a Fourier transform $\boldsymbol F$ are  exploited on the auto-correlation vector $\boldsymbol a$ of the sensed signals. Since using smoothing operator, the PSD is almost piecewise constant generally. Therefore the number of significant non-zero values in the edge vector of PSD $\boldsymbol z$ is so sparse. The edge vector can be found by using $\boldsymbol z = \Gamma  \boldsymbol s$ where in the most simple case $ \Gamma$ is a differential operator as expressed in \cite{CrCs3}. The dictionary of the spare representation ($\boldsymbol D$) can be found as $\boldsymbol z = \Gamma  \boldsymbol W \boldsymbol F \boldsymbol a = \boldsymbol D^{-1} \boldsymbol a $. Now, reconstructing the sparse edge vector of the PSD is possible from the sensed measurements $\boldsymbol y = \Phi  \boldsymbol a$. Finally, the estimated PSD, $\hat{\boldsymbol s}$, can be achieved by $\hat{\boldsymbol s} = \boldsymbol G \hat{\boldsymbol z}$ where $\hat{\boldsymbol z}$ is the estimated edge vector and $\boldsymbol G $ is the cumulative sum matrix (i.e. a lower triangular matrix with $+1$ elements). 

Suppose that $M^2$ sensors should capture a PSD $\boldsymbol s_j \in R^N$ ($j \in {1,2,..., M^2}$) and are distributed in the area. It can be expected that there is a shared common component $\boldsymbol s_c \in R^N$ between the $J$  neighbor sensors which constitute a Group of Sensors (GoS), such that $\boldsymbol s_j=\boldsymbol s_c + \boldsymbol s_{{inn}_j}$. $\boldsymbol s_{{inn}_j} \in R^N$ is the innovation part of each PSD $\boldsymbol s_j$. PSDs can be represented as a cumulative sum of their edge vector as $\boldsymbol s_j= \boldsymbol G \boldsymbol z_{j}$ by matrix $\boldsymbol G \in R^{N\times N}$. Obviously, we have sparse $\boldsymbol z_c$ and $\boldsymbol z_{{inn}_j}$s which belong to space $R^N$ with different sparsity levels. Therefore, $\boldsymbol s_j=\boldsymbol G \boldsymbol z_{j}=\boldsymbol G (\boldsymbol z_c+\boldsymbol z_{{inn}_j})$ and/or $\boldsymbol s_c=\boldsymbol G \boldsymbol z_c, \boldsymbol s_{{inn}_j}=\boldsymbol G \boldsymbol z_{{inn}_j}$. $\Phi_j \in R^{w_j \times N}$ is an individual measurement matrix for the $j$th sensor and its sensed measurements $\boldsymbol y_j = \Phi_j \boldsymbol a_j$ should be sent to the FC. $\boldsymbol r_j \in R^{w_j}$ is the received signal by the FC according to the sent $\boldsymbol y_j$. Notice that in the rest of the paper the $\hat{}$ denotes the reconstructed vectors or signals. 

In this document, we seek to propose criteria for compressing the sensed PSDs of the sensors to reduce the amount of the transmitted data from sensors to FC and also deal with the channel imperfection transmission.



\section{Proposed Approach}

The the simplest idea to reconstruct the PSD $\boldsymbol s_j$ is $\hat{\boldsymbol s}_j=  \boldsymbol G \hat{\boldsymbol z}_j$ where $\hat{\boldsymbol z_j}$ is attained by solving the BPDN \cite{BPDN} inspired problem eq. \eqref{eq.00}. Also we can use the shared common component $\boldsymbol s_c$ in a GoS based on the Joint Sparsity model (JSM) \cite{DCS} and therefore, reconstruct the data in lower measuring (sensing) rate. Inspired from JSM, equations (\ref{eq.10}) to (\ref{eq.13}) are defined to model recontsruction for a $J$ neighbor sensors (a GoS). Therefore, the desired PSDs can be yielded by $\hat{\boldsymbol s}_j=\boldsymbol G (\hat {\boldsymbol z}_c + \hat{{\boldsymbol z}}_{{inn}_j})$ where $\hat {\boldsymbol z}_c$ and $\hat{{\boldsymbol z}}_{{inn}_j}$s are located in the found $\hat{\boldsymbol z}  = {\left[ {\begin{array}{*{20}{c}} {{\hat{\boldsymbol z} _c}^T}&{{{\hat{\boldsymbol z}}_{{inn}_1}}^T}& \cdots &{{{\hat{\boldsymbol z}}_{{inn}_J}}^T} \end{array}} \right]^T}$ vector. $\hat{\boldsymbol z}$ is computed by solving the optimization problem in equation (\ref{eq.14}) where $\boldsymbol r \in R^{W}$, $\boldsymbol n \in R^{W}$, $\boldsymbol z \in R^{N(J+1)}$, $\Psi \in R^{W \times N(J+1)}$ and $W = \sum\limits_{j = 1}^J {{w_j}}$.
\begin{align}
\hat{\boldsymbol z}_j= \mathop {\min }\limits_{\acute{\boldsymbol z}_j}   \frac{1}{2} \left\| {\boldsymbol r_j- \Phi_j D \acute{\boldsymbol z}_j\|} \right._2^2 +  \lambda \left\|\acute{\boldsymbol z}_j \right\|_1. \label{eq.00} \ \ \ \ \ \ \ \\
\boldsymbol r= \Psi \boldsymbol z + \boldsymbol n \label{eq.10} \ \ \ \ \ \ \ \ \ \  \ \ \ \ \ \ \ \ \ \ \ \ \\
\boldsymbol r = \left[ {\begin{array}{*{20}{c}}  
{{\boldsymbol r_1}^T} \ 
{{\boldsymbol r_2}^T} \ 
 \hdots \ 
{{\boldsymbol r_J}^T}
\end{array}} \right]^T \ \ \ \ \ \ \ \ \ \ \ \nonumber \\ \boldsymbol n = \left[ {\begin{array}{*{20}{c}}
{{\boldsymbol n_1}^T} \ 
{{\boldsymbol n_2}^T} \ 
 \hdots \ 
{{\boldsymbol n_J}^T}
\end{array}} \right]^T \ \ \ \ \ \ \ \ \ \ \label{eq.11} \\  \boldsymbol z = \left[ {\begin{array}{*{20}{c}}
{{\boldsymbol z_c}^T} \ 
{{\boldsymbol z_{{inn}_1}}^T} \ 
 \hdots \ 
{{\boldsymbol z_{{inn}_J}}^T} \ 
\end{array}} \right]^T \ \ \ \ \ \ \ \nonumber \\
\Psi  = \left[ {\begin{array}{*{20}{c}}
{\Phi _1} \boldsymbol D& {\Phi _1} \boldsymbol D&{\begin{array}{*{20}{c}}
0& \cdots 
\end{array}}&0\\
{\Phi _2} \boldsymbol D&0&{\begin{array}{*{20}{c}}
{\Phi _2} \boldsymbol D& \cdots 
\end{array}}& \vdots \\
 \vdots &{}&{\begin{array}{*{20}{c}}
{}& \ddots 
\end{array}}&0\\
{\Phi _J}\boldsymbol D&0&{\begin{array}{*{20}{c}}
 \cdots 
\end{array}}&{\Phi _J}\boldsymbol D
\end{array}} \right] \label{eq.13} \\
\hat{\boldsymbol z}= \mathop {\min }\limits_{\acute{\boldsymbol z}}   \frac{1}{2} \left\| {\boldsymbol r - \Psi \acute{\boldsymbol z}\|} \right._2^2 +  \lambda \left\|\acute{\boldsymbol z}\right\|_1  \ \ \ \ \ \ \ \ \label{eq.14}
\end{align}

But, assume a scenario in which the common part $\boldsymbol z_c$ is known by the FC and fix for a while of time. For example, the holding time is 30 seconds while the network is sensed in each 0.3 seconds. Here, in order to enhance the model, we remove the common part from the reconstruction equation. Equation (\ref{eq.10}) can be rewritten in form of equation (\ref{eq.16}) and split by using a combination of two distinct parts: Common part $\boldsymbol z_c$ and Innovation part $\boldsymbol z_I$. The reconstruction formula can be modified to just find the innovation parts of the PSDs by \eqref{eq.17} where $\boldsymbol r_{inn}=\boldsymbol r-\boldsymbol A \boldsymbol z_c$. Consequently, the PSD of each sensor will be found by $\hat{\boldsymbol s}_j= \boldsymbol G (\boldsymbol z_c + \hat{{\boldsymbol z}}_{{inn}_j})$ where $\hat{{\boldsymbol z}}_{{inn}_j}$s are located in the computed $\hat{{\boldsymbol z}}_I$ vector. This modification brings faster solution and also better reconstruction accuracy. 
\begin{align}
\boldsymbol r=\left[ {\boldsymbol A\| \boldsymbol H} \right] \left[ {\begin{array}{*{20}{c}}
{{\boldsymbol z_c}^T} \ 
{{\|}} \ 
{{\boldsymbol z_{{inn}_1}}^T} \
 \hdots \ 
{{\boldsymbol z_{{inn}_J}}^T}
\end{array}} \right]^T + \boldsymbol n \nonumber \\ = \boldsymbol A \boldsymbol z_c +\boldsymbol H \boldsymbol z_I +\boldsymbol n  \label{eq.16} \ \ \ \ \ \ \ \ \ \ \ \ \\ 
\hat{{\boldsymbol z}}_I= \mathop {\min }\limits_{\acute{\boldsymbol z_I}}   \frac{1}{2} \left\| {\boldsymbol r_{inn} - \boldsymbol H \acute{\boldsymbol z_I}\|} \right._2^2 +  \lambda \left\|\acute{\boldsymbol z_I}\right\|_1 \label{eq.17} \ \ \ \
\end{align} 

Now we try to find the optimum common component $\boldsymbol z_c$ labeled as $\boldsymbol z_{c_{opt}}$. Since our proposed model is based on JSM, the first well known approach to find the optimum is solving the JSM based optimization problem \eqref{eq.19} where $\overline {\boldsymbol G}$ is a matrix constructed by arranging $\boldsymbol G$s similar to eq. \eqref{eq.13} ($\phi_j$s are replaced by $G$s). But remember that the desired variables in the proposed eq. (\ref{eq.17}) are just innovation parts $\boldsymbol z_I$ and the optimization constraint is just the maximum sparsity of the innovation parts. Therefore the problem to find the $\boldsymbol z_{c_{opt}}$ can be exchanged to \eqref{eq.20} where the $\boldsymbol z_{c_{opt}}$ is a part of the found  $\boldsymbol z_{opt}$ vector. The more details and achievements of these proposed criteria are published in \cite{Mine}.
\begin{align}
\label{eq.18}
\boldsymbol z_{opt} = \left[ {\begin{array}{*{20}{c}}
{{\boldsymbol z_{c_{opt}}}^T} \ 
{{\boldsymbol z_{{inn}_{1_{opt}}}}^T} \ 
 \hdots \ 
{{\boldsymbol z_{{inn}_{J_{opt}}}}^T}
\end{array}} \right]^T \nonumber \\ \boldsymbol s_{all} = \left[ {\begin{array}{*{20}{c}}
{{\boldsymbol s_1}^T} \ 
{{\boldsymbol s_2}^T} \ 
 \hdots \
{{\boldsymbol s_J}^T} 
\end{array}} \right]^T \ \ \ \ \ \ \ \ \ \  \\
\boldsymbol z_{opt}= \mathop {\min }\limits_{\acute{\boldsymbol z}_{opt}} \left\|\acute{\boldsymbol z}_{opt}\right\|_{1} subject \ to \  \boldsymbol s_{all} =  \overline {\boldsymbol G} \acute{\boldsymbol z}_{opt}  \label{eq.19} \\
\boldsymbol z_{opt}= \mathop {\min }\limits_{\acute{\boldsymbol z}_{opt}} \sum\limits_{j = 1}^J \left|{\acute{\boldsymbol z}_{{inn}_{j_{opt}}}}\right|_1 subject \ to \  \boldsymbol s_{all} =  \overline {\boldsymbol G}  \acute{\boldsymbol z}_{opt} \label{eq.20}
\end{align}

Now lets use the known and fixed for a while common part $\boldsymbol z_{opt}$ more. The above mentioned criteria was based on our model in \eqref{eq.10}. But, it is clear that, recovering the PSDs by using eq. (\ref{eq.00}) and also eq. (\ref{eq.14}) only can be useful when there is no signifant difference between $\boldsymbol r_j$ and $\boldsymbol y_j$. In order to embed the effect of this imperfection in the model, we have proposed to estimate a destructive filter $\boldsymbol \beta_j \in R^{w_j}$ and exploit it with the model by circular convolution as $\boldsymbol r_j=\boldsymbol y_j \odot \boldsymbol\beta_j+\boldsymbol n_j$. Similarly we can model the received signal of the common part $\boldsymbol r_{c_j}$ as $\boldsymbol r_{c_j}=\boldsymbol y_{c_j} \odot \boldsymbol\beta_j +\boldsymbol \breve{n}_j$ or in the matrix multiplication form as $\boldsymbol r_{c_j}= {\mathop {\boldsymbol Y}\limits^o}_{c_j} \boldsymbol\beta_j + \boldsymbol \breve{n}_j$ where ${\mathop {\boldsymbol Y}\limits^o}_{c_j} \in R^{w_j \times w_j}$ is the circulant matrix of $\boldsymbol y_{c_j} \in R^{w_j}$. Reformulate eq. (\ref{eq.10}) to equation (\eqref{eq.14.7}) brings new model where $\overline{\overline {\boldsymbol B}} \in R^{W \times W}$ is defined in (\ref{eq.14.8}) and ${\mathop {\boldsymbol B}\limits^o}_{j} \in R^{(w_j)\times w_j}$ is the circulant matrix of the $\hat{\boldsymbol\beta_j}$. Therefore, estimated  impulse response of the destructive filter can be achieved by solving the optimization problem $\hat{\boldsymbol\beta}_j= \mathop {\min }\limits_{\acute{\boldsymbol\beta_j}} \left\| \boldsymbol r_{c_j}-{\mathop {\boldsymbol Y}\limits^o}_{c_j} \acute{\boldsymbol\beta_j} \right\|_2$. The Reconstruction criterion for this new model is similar to eq. \eqref{eq.17} while the entities of matrix $\boldsymbol H$ are modified.

\begin{align}
\boldsymbol r=\overline{\overline {\boldsymbol B}}  \Phi \Psi \boldsymbol z + \boldsymbol n \label{eq.14.7}  \ \ \ \ \ \ \ \\
\overline{\overline {\boldsymbol B}}  = \left[ {\begin{array}{*{20}{c}}
{{\mathop {\boldsymbol B}\limits^o}_1 }&0& \cdots &0\\
0&{{\mathop {\boldsymbol B}\limits^o}_2 }& \cdots &0\\
 \vdots & \cdots & \ddots & \vdots \\
0& \cdots &0&{{\mathop {\boldsymbol B}\limits^o}_J}
\end{array}} \right] \label{eq.14.8}
\end{align}


\section{Experiments}
\begin{figure}
  \centering
  \includegraphics[width= .4 \textwidth]{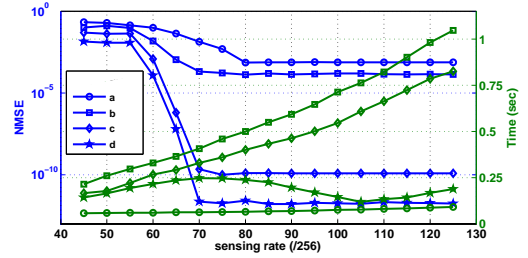}
  \caption{Comparison between compression ability and reconstruction time of the mentioned models. a) Individual reconstruction, eq. (1). b) JSM reconstruction, eq. (5). c) Proposed reconstruction eq. (7) with $\boldsymbol z_{c_{opt}}$ found by eq. (9). d) Proposed reconstruction eq. (7) with $\boldsymbol z_{c_{opt}}$ found by eq. (10) \cite{Mine}}.
  \label{MseIEEE}
\end{figure}

\begin{figure}
  \centering
  \includegraphics[width= .43 \textwidth]{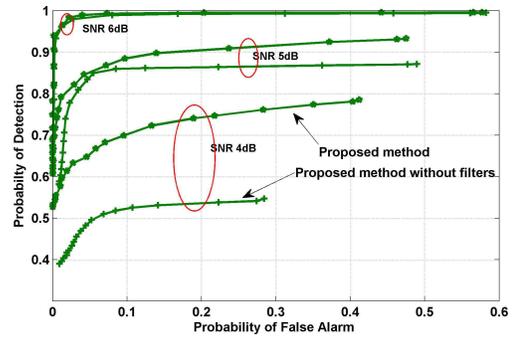}
  \caption{Improvements caused by using and computing disturbance filters in the model for lossy networks.}
  \label{RocBER}
\end{figure}

To validate our algorithm, a sensor network is considered in which has $M  \times M (M=12)$ uniformly distributed sensors with $J=4$ in each GoS. The details of the case study is included in \cite{Mine}. Fig. \ref{MseIEEE} shows the achievements of the proposed method with respect to reducing sensing rate and reconstruction time in a loss-less network. Fig. \ref{RocBER} shows the effect of embedding the disturbance filters in the reconstruction criterion.

\section{Discussion and Conclusions}
Since the CR sensors are distributed in the region of support, in order to construct the PSD maps, the sensed PSD by each sensor should be transmitted to a FC. Therefore, when the number of sensors is large, transmitting this type of overhead data can be challenging. In this paper some criteria are proposed based on using shared part of the signals to compress the PSDs and  reconstruct them more robustly. 


\begin{thebibliography}{1}

\bibitem{CrCs3}
D. Sundman et al., ``On the use of compressive sampling for wide-band spectrum sensing," {\it IEEE - ISSPIT}, 2010.

\bibitem{BPDN}
P.R. Gill et al., ``The In-Crowd Algorithm for Fast Basis Pursuit Denoising," {\it IEEE Trans. Signal Process.}, vol. 59, no. 10, 2011. 

\bibitem{DCS}
D. Baron et al., (2009, jan.) ``Distributed Compressed Sensing," [online] Availabe: http://arxiv.org/abs/0901.3403/ 2006.

\bibitem{Mine}
M. Eslami et al., ``A Centralized PSD Map Construction by Distributed Compressive Sensing," {\it IEEE Communications Letters}, vol. 19,  no. 3, 2015.

\end{thebibliography}
\end{document}